\documentclass[10pt, a4paper]{article}
\usepackage{lrec}
\usepackage{graphicx}
\usepackage{amsmath}
\usepackage{multirow}
\usepackage{url}
\usepackage{booktabs}
\usepackage[normalem,normalem,normalbf]{ulem}
\usepackage{tabularx}
\usepackage{soul}
\usepackage{tikz}
\usepackage[latin1]{inputenc}
\usepackage{xstring}
\usepackage{tikz}
\def\checkmark{\tikz\fill[scale=0.8](0,.35) -- (.25,0) -- (1,.7) -- (.25,.15) -- cycle;}
\definecolor{FdWgreen}{rgb}{0.0, 0.42, 0.24}

\title{\textbf{Semi-supervised acoustic modelling for five-lingual code-switched ASR using automatically-segmented soap opera speech}}

\name{N. Wilkinson$^1$, A. Biswas$^1$,  E. Y\i lmaz$^2$, F. de Wet$^1$, E. van der Westhuizen$^1$, T.R. Niesler$^{1}$}

\address{$^1$  Department of Electrical and Electronic Engineering, Stellenbosch University, South Africa \\
         $^2$ Department of Electrical and Computer Engineering, National University of Singapore, Singapore \\
         \{nwilkinson, abiswas, fdw, ewaldvdw, trn\}@sun.ac.za, emre@nus.edu.sg}
         
\abstract{
This paper considers the impact of automatic segmentation on the fully-automatic, semi-supervised training of automatic speech recognition (ASR) systems for five-lingual code-switched (CS) speech. 
Four automatic segmentation techniques were evaluated in terms of the recognition performance of an ASR system trained on the resulting segments in a semi-supervised manner.
The systems' output was compared with the recognition rates achieved by a semi-supervised system trained on manually assigned segments. 
Three of the automatic techniques use a newly proposed convolutional neural network (CNN) model for framewise classification, and include a novel form of HMM smoothing of the CNN outputs.
Automatic segmentation was applied in combination with automatic speaker diarization. The best-performing segmentation technique was also tested without speaker diarization.
An evaluation based on 248 unsegmented soap opera episodes indicated that voice activity detection (VAD) based on a CNN followed by Gaussian mixture model-hidden Markov model smoothing (CNN-GMM-HMM) yields the best ASR performance. The semi-supervised system trained with the resulting segments achieved an overall WER improvement of 1.1\% absolute over the system trained with manually created segments.
Furthermore, we found that system performance improved even further when the automatic segmentation was used in conjunction with speaker diarization.
\newline 
}

\begin{document}

\maketitleabstract
 
\section{Introduction}
\label{sec:intro}
Code-switching is the alternation between two or more languages by a single speaker during discourse, and is a common phenomenon in multilingual societies.
In South Africa, for example, 11 official and geographically co-located languages are in use, including English which serves as the lingua franca.
Here, speakers frequently code-switch between English, a highly-resourced language, and their Bantu mother tongue, which is in comparison highly under-resourced.

The automatic recognition of code-switched speech has become a topic of growing research interest, as reflected by the increasing number of language pairs that have recently been studied.
While English-Mandarin has received extensive attention~\cite{li2013improved,zeng2018end,vu2012first,taneja2019exploiting}, other language pairs such as Frisian-Dutch \cite{yilmaz2016investigating,yilmaz2018semi}, Hindi-English \cite{pandey2018phonetically,emond2018transliteration,ganji2019iitg}, English-Malay \cite{ahmed2012automatic}, Japanese-English \cite{nakayama2018speech} and French-Arabic \cite{amazouz2019addressing} have also attracted interest.
We have introduced the first South African corpus of multilingual code-switched soap opera speech in \cite{van2018city}. 

For code-switched speech, the development of robust acoustic and language models that are able to extend across language switches is a challenging task.
When one or more of the languages are under-resourced, as it is in our case, data sparsity limits modelling capacity and this challenge is amplified.
Acoustic data that includes code-switching is extremely hard to find, because it usually does not occur in formal conversation, such as broadcast news, and also because it requires skilled multilingual language practitioners for its annotation.
The result is that manually-prepared datasets including code-switched speech in Africa are destined to remain rare and small.
 
In previous work, we have demonstrated that multilingual training using in-domain soap opera code-switched speech and poorly matched monolingual South African speech improves the performance of both bilingual and five-lingual automatic speech recognition (ASR) systems when the additional training data is from a closely-related language~\cite{biswas2018IS,biswas2018improving}. 
Specifically, isiZulu, isiXhosa, Sesotho and Setswana belong to the same Bantu language family and were found to complement each other when combined into a multilingual training set for acoustic modelling.  
%It has been also found that better performance is achieved when multilingual acoustic models are trained on more in-domain data compared to the same amount of out-of-domain data \cite{biswas2018improving}.
Hence, increasing the amount of in-domain code-switched speech data is a reliable way to achieve more robust ASR.
%However, it is been a difficult and time consuming task to compile in-domain code-switch data from multilingual soap opera speech as it involves skilled human annotators and transcribers.
However, the development of such in-domain data is a time-consuming and costly endeavour as it requires highly skilled human annotators and transcribers.

To address this lack of annotated data, automatically transcribed training material has been shown to be useful in under-resourced scenarios using semi-supervised training ~\cite{thomas2013deep,yilmaz2018semi,Guo2018}.
This strategy was successfully implemented on South African code-switched speech to obtain bilingual and five-lingual ASR systems using 11.5 hours of manually segmented but untranscribed soap opera speech~\cite{biswas2019IS2}.
Recently a study has analyzed the performance of batch-wise semi-supervised training on South African code-switched ASR \cite{biswas2020LREC}.
%Here the semi-supervised bilingual systems trained with automatic transcriptions generated by the five-lingual transcription system achieved the best performance.
However, manual segmentation of the raw soap opera audio by skilled annotators was still required to identify the speech that is useful for ASR.
Therefore, this approach is not fully automatic which remains an impediment in resource-scare settings.

%\todo{Ewald: Is this reference to radio broadcasts correct?} 
In this study, we apply four automated approaches to the segmentation of soap opera speech and investigate the effect on ASR performance.
A conventional energy-based voiced activity detector (VAD) \cite{povey2011kaldi}, as well as CNN-HHM and CNN-GMM-HMM systems that we have developed are used to distinguish between speech, music and noise.
In addition, an X-vector DNN embedding system is used for speaker diarization to obtain speaker specific metadata for three of the segmentation approaches~\cite{snyder2018x}.
%To make the system fully automatic and speed up the development of robust South-African code-switch ASR by deploying automatic segmentation technique followed by speaker diarisation.
%Four different automatic transcriptions were considered for this study.
For the experiments, 248 complete soap opera episodes, each approximately 22 minutes in length, were used.
It is important to note that we also have the manual segmentation (approximately 24 hours of speech) of these 248 episodes and can therefore perform a comparative evaluation with the automated approaches.
Semi-supervised systems trained using the manually-segmented speech were used as baselines and compared with systems trained on speech identified by the automatic approaches.
 
Pseudo-labels or transcriptions of automatically segmented speech were generated using our best baseline systems trained on 21 hours manually transcribed speech and 11 hours of manually segmented but automatically transcribed speech.
Given the multilingual nature of the data, the transcription systems must not only provide the orthography, but also the language(s) present at each location in each segment.
To achieve this, each segment was presented to four individual code-switching systems as well as to a five-lingual system.

%\todo{Ewald: Removed BNF text.}
%Bottleneck features, and in particular multilingual bottleneck features, have been shown to outperform traditional MFCC and PLP features in many spoken language processing tasks \cite{Vesely2012,fer2017multilingually,hermann2018multilingual}.
%To the best of our knowledge, we developed the first multilingual bottleneck feature (BNF) extractor trained on nine indigenous South African Bantu languages.
%It has been shown that the better-resourced monolingual speech can contribute to a multilingual code-switch ASR trained on soap opera speech \cite{biswas2019journal}.
%However, the improvement is not so significant considering using more than an order of magnitude monolingual out-of-domain speech and acoustic model training is quite slow due to computational constraints.
%Thus, training a BNF extractor with monolingual speech and using bottleneck embedding with traditional MFCC acoustic features is a relevant option.  

% ----------------------------------------
\section{Data}
% ----------------------------------------
%\subsection{Multilingual soap opera data}
\label{SEC:corpus}
For experimentation, we use a corpus of multilingual, code-switched speech compiled from South African soap opera episodes.
This corpus contains both manually and automatically-annotated speech divided into four language pairs: English-isiZulu (EZ), English-isiXhosa (EX), English-Setswana (ET), and English-Sesotho (ES).
Of the Bantu languages, isiZulu and isiXhosa belong to the Nguni language family while Setswana and Sesotho are Sotho-Tswana languages. 

The corpus contains 8\,275, 11\,352, 6\,169, 1\,902 and 2\,792 unique English, isiZulu, isiXhosa, Setswana and Sesotho words, respectively.
IsiZulu and isiXhosa have relatively large vocabularies due their agglutinative nature and conjunctive writing system.
Although Setswana and Sesotho are also agglutinative, they use disjunctive writing systems which result in smaller vocabularies than isiZulu and isiXhosa.
The speech in the soap opera episodes is also typically fast and often expresses emotion. 
These aspects of the data in combination with the high prevalence of code-switching makes it a challenging corpus for conducting ASR experiments. 

\subsection{Manually segmented and transcribed data (ManT)}
\label{SEC:corpus:transcribed}
Our first code-switching ASR systems were developed and evaluated on 14.3 hours of speech divided into four language-balanced sets, as described in~\cite{van2018city}. 
In addition to the language-balanced sets, approximately another nine hours of manually transcribed speech was available.
This additional data is dominated by English and was initially excluded from our training set to avoid bias.
However, pilot experiments indicated that, counter to expectations, its inclusion enhanced recognition performance in all languages.
The additional data was therefore merged with the balanced sets for the experiments described here. 
Of this, 21.1 hours is used as a training set, 48 minutes as a development set, and 1.3 hours as a test set.
The composition of the unbalanced training set is shown in Table~\ref{tab:duration_unbalanaced_corpora}.

\begin{table}[h]
\centering 
%\scriptsize
\footnotesize
\renewcommand*{\arraystretch}{1.0}
%\resizebox{\columnwidth}{!}{
\begin{tabular*}{0.47\textwidth}{@{\extracolsep{\fill}} l r @{\hspace{1em}} r @{\hspace{1em}} r @{\hspace{1em}} r @{\hspace{1em}} r @{\hspace{1em}} r @{}}
\toprule
{\bf{Language}} & \begin{tabular}[c]{@{}c@{}}\bf{Mono}\\  (m)\end{tabular} & \begin{tabular}[c]{@{}c@{}}\bf{CS}\\ (m)\end{tabular}&
\begin{tabular}[c]{@{}c@{}}\bf{Total}\\ (h)\end{tabular}&
\begin{tabular}[c]{@{}c@{}}\bf{Total}\\ (\%)\end{tabular}&
\begin{tabular}[c]{@{}c@{}}\textbf{Word}\\ \textbf{tokens}\end{tabular} & \begin{tabular}[c]{@{}c@{}}\textbf{Word} \\ \textbf{types}\end{tabular} \\ \midrule
English & 755.0              & 121.8  & 14.6 & 69.3 & 194\,426              & 7\,908              \\
isiZulu & 92.8               & 57.4  & 2.5 & 11.9 & 24\,412              & 6\,789              \\
isiXhosa & 65.1               & 23.8  & 1.5 & 7.0  & 13\,825              & 5\,630              \\ 
Setswana & 36.9               & 34.5  & 1.2 & 5.6  & 21\,409              & 1\,525              \\
Sesotho & 44.7               & 34.0  & 1.3 & 6.2  & 22\,226              & 2\,321              \\
\midrule
{\bf{Total}} & 994.5              & 271.5  & 21.1 & 100.0  & 276\,290              & 24\,170              \\ \bottomrule
\end{tabular*}%
%}
\caption{Duration in minutes (m) and hours (h) as well as word type and token counts for the unbalanced training set.}
\label{tab:duration_unbalanaced_corpora}
\end{table}

An overview of the composition of the development (Dev) and test (Test) sets for each language pair is given in Table \ref{tab:corpora_stat}.
The table includes values for the total duration as well as the duration of the monolingual  and code-switched  segments. 
The test sets contain no monolingual data and a total of approximately 4\,000 language switches (English-to-Bantu and Bantu-to-English).

\begin{table}[h]
\small
	\centering
    \renewcommand*{\arraystretch}{0.9}
		\begin{tabular*}{0.47\textwidth}{@{\extracolsep{\fill}}c r r r r r @{}}
			\toprule
			\multicolumn{6}{c}{\textbf{English-isiZulu}} \\
			%\midrule
			& emdur & zmdur & ecdur & zcdur & \textbf{Total} \\
		%	\textbf{Train} & 1.55h & 1.55h & 45.86 & 56.99 & 4.81h \\ \hline
			\textbf{Dev} & 0.0 & 0.0 & 4.0 & 4.0 & 8.0 \\
			\textbf{Test} & 0.0 & 0.0 & 12.8 & 17.9 & 30.4 \\ \midrule
		%	\textbf{Total} & 1.55h & 1.55h & 1.04h & 1.31h & 5.45h \\ \hline
			\multicolumn{6}{c}{\textbf{English-isiXhosa}} \\
			%\midrule
			& emdur & xmdur & ecdur & xcdur & \textbf{Total} \\
		%	\textbf{Train} & 65.22m & 53.55m & 18.04m & 23.73m & 160.54m \\ \hline
			\textbf{Dev} & 2.9 & 6.5 & 2.2 & 2.1 & 13.7 \\
			\textbf{Test} & 0.0 & 0.0 & 5.6 & 8.8 & 14.3 \\ \midrule
		%	\textbf{Total} & 68.08m & 60.03m & 25.81m & 34.64m & 3.143h \\ \hline
			\multicolumn{6}{c}{\textbf{English-Setswana}} \\
			%\midrule
			 & emdur & tmdur & ecdur & tcdur & \textbf{Total} \\
		%	\textbf{Train} & 40.4m & 30.96m & 34.37m & 34.01m & 139.74m \\ \hline
			\textbf{Dev} & 0.8 & 4.3 & 4.5 & 4.3 & 13.8 \\
			\textbf{Test} & 0.0 & 0.0 & 8.9 & 9.0 & 17.8 \\ \midrule
		%	\textbf{Total} & 41.16m & 35.22m & 47.78m & 47.24m & 2.86h \\ \hline
			\multicolumn{6}{c}{\textbf{English-Sesotho}} \\
			%\midrule
			 & emdur & smdur & ecdur & scdur & \textbf{Total} \\
		%	\textbf{Train} & 49.34m & 35.32m & 23.02m & 34.04m & \multicolumn{1}{l|}{141.72m} \\ \hline
			\textbf{Dev} & 1.1 & 5.1 & 3.0 & 3.6 & 12.8 \\
			\textbf{Test} & 0.0 & 0.0 & 7.8 & 7.7 & 15.5 \\ \bottomrule
		%	\textbf{Total} & 50.43m & 40.37m & 33.85m & 45.37m & \multicolumn{1}{l|}{2.83h} \\ \hline
		\end{tabular*}%
	\caption{Duration (minutes) of English, isiZulu, isiXhosa, Sesotho, Setswana monolingual (mdur) and code-switched (cdur) segments in the code-switching development and test sets.}
	\label{tab:corpora_stat}

\end{table}

\subsection{Manually segmented automatically transcribed data: Expert segmentation (AutoT$_{\rm Exp}$)}
\label{SEC:corpus:untranscribed1}
During corpus development, approximately 11 hours of manually segmented speech (representing 127 different speakers) was produced in addition to the manually transcribed data described in the previous section.
Segmentation was performed manually by experienced language practitioners.
This dataset (AutoT$_{\rm Exp}$) was automatically transcribed during our initial investigations into semi-supervised acoustic model training, resulting in 7\,951 EZ, 3\,796 EX, 11\,415 ES and 128 ET segments~\cite{biswas2019IS2}.

\subsection{Manually segmented automatically transcribed data: Non-expert segmentation (AutoT$_{\rm NonE}$)}
\label{SEC:corpus:untranscribed2}
A subsequent phase of corpus development, currently still underway, has produced manual segmentations for a further 248 soap opera episodes.
These 248 episodes amount to 89 hours of audio data before segmentation, and 23 hours of speech data (AutoT$_{\rm NonE}$) after segmentation.
The segmentation was not performed by language experts and is therefore expected to be less accurate than that of the AutoT$_{\rm Exp}$ data.
Furthermore South African languages other than the five present in the transcribed data are known to occur in this batch, but to a limited extent.

This set of 248 episodes was used in the automatic segmentation experiments described in the next section because the manually assigned segment labels were available as a reference in the form of AutoT$_{\rm NonE}$.

\section{Automatic Segmentation}
\label{sec:auto_seg}
%\textcolor{red}{\textbf{NICK}}
%V1: First one (Energy based VAD folowed by Spk dia)
%V2: CNN-HMM-Spk Dia
%V3: Improved CNN-HMM-Spk Dia (I guess so)
%V4: CNN-HMM (Without spk Dia)

A number of automatic segmentation techniques were considered as alternatives to the labour-intensive process of manually segmenting the soap operas.
Different voice activity detection (VAD) approaches were combined with the X-vector DNN embedding-based speaker diarization system introduced in~\cite{snyder2018x} to obtain speaker labels.
In subsequent ASR experiments, the best performing VAD technique was also evaluated without speaker diarization.

\subsection{VAD${_1}$: Energy-based}
In our first experiment, the X-vector diarization recipe provided in the Kaldi toolkit was applied using an X-vector DNN model pre-trained on wide-band VoxCeleb data~\cite{povey2011kaldi,Nagrani17vox,Chung18bvox}.
This system uses 24-dimensional filterbank features based on 25ms frames.
Speech frames are identified using a simple energy threshold and are subsequently passed to the pre-trained DNN which extracts the X-vectors.
Finally, probabilistic linear discriminant analysis (PLDA) is applied to the X-vectors, and agglomerative hierarchical clustering is used to assign speaker labels.

A difficulty observed when using this approach was that, while a simple energy VAD works reasonably well under low noise conditions where most frames are speech, it performs poorly when confronted with our soap opera data in which extensive non-speech segments containing music and other sounds are common.
Post-diarization listening tests revealed that many non-speech segments were still present in the data classified as speech.
Adjustment of the VAD threshold to more aggressively remove non-speech segments resulted in the loss of many speech segments.
%Adjustment of the VAD threshold to more aggressively remove non-speech segments resulted in the loss of many speech segments, as shown in Table \ref{tab:VADres}.

\begin{table}
\resizebox{\columnwidth}{!}{
\begin{tabular}{@{}lcc@{}}
\toprule
\textbf{Layer}          & \textbf{Kernels/Nodes}        & \textbf{Activation Function}  \\ \midrule
Convolutional\_1        & 32 (3x3 kernel)               & ReLU                          \\
Max\_Pooling\_2         & -                             & -                             \\
Convolutional\_3        & 64 (3x3 kernel)               & ReLU                          \\
Max\_Pooling\_4         & -                             & -                             \\
Convolutional\_5        & 64 (3x3 kernel)               & ReLU                          \\
Flatten\_6              & 1024                          & -                             \\
Fully\_Connected\_7     & 64                            & ReLU                          \\
Fully\_Connected\_8     & 2                             & Sigmoid                       \\\bottomrule
\end{tabular}
}
\caption{The CNN architecture used in the VAD systems.}
\label{tab:CNNarc}
\end{table}

\subsection{VAD${_2}$: CNN-HMM}
At the time of writing, the X-vector based system achieved state-of-the-art performance in diarization tasks. 
However, the energy based VAD it uses limits performance.
For this reason, efforts to improve automatic segmentation focused on developing improved VAD.
Recently, CNNs have been successfully applied to the task of VAD \cite{ThomasCNNVAD}, and both large and small architectures have been found to perform well \cite{SehgalCNNVAD,HersheyCNNVAD}.
In our resource-constrained setting, computational efficiency is important since VAD will most likely occur on a mobile device.

We introduce a small CNN architecture ($\approx$120 000 parameters) implemented in Python, using Tensorflow (v2.0.0) and Keras (v2.2.4-tf), to create a fast, lightweight VAD system whose architecture is shown in Table \ref{tab:CNNarc}. 
This system computes 32-dimensional log-mel filterbank energies using a frame length of 10ms and then stacks these over 320ms to form 32x32 spectrogram features as input to the CNN.
The CNN was trained on the balanced subset ($\approx$ 53 hours) of Audio Set \cite{GemmekeAudioSet} to classify frames as containing ``speech'' and/or ``non-speech''.

%A 50\% split of the AVA-Speech dataset \cite{ChaudhuriAVA} was used as a test set.
%The remaining 50\% was used as a training set for the HMM smoothing models discussed below.
%The dataset contains $\approx$46 hours of densely labeled, multilingual movie data, with the following class labels: ``NoSpeech'',``CleanSpeech'', ``Speech+Music'' and ``Speech+Noise''.
%This dataset provides similar conditions to our target domain of soap opera data, making it a suitable test set.
%Furthermore, it is accompanied by baseline results, tested on the WebRTC project VAD \cite{WebRTC}, and two CNN systems based on \cite{HersheyCNNVAD}.
%The smaller of the two, \textit{tiny320}, is similar in size to our CNN, also containing three convolutional layers, whilst \textit{resnet960} is based on the large ResNet-50 architecture \cite{resnet50}.
%Results for all VAD systems are shown in Table \ref{tab:VADres}.

While our CNN on its own performs well at the VAD task, it fails to capture temporal patterns in the data and was observed to often mislabel single frames within extended sections of speech or non-speech.
In an initial attempt to address this, we introduce a HMM for smoothing.
The AVA-Speech dataset \cite{ChaudhuriAVA} was used to train our HMMs, and for testing of the final VAD.
The full dataset contains $\approx$~46 hours of densely labeled, multilingual movie data, with the following class labels: ``NoSpeech'',``CleanSpeech'', ``Speech+Music'' and ``Speech+Noise''.
AVA-Speech (train), a randomly selected $\approx$~23 hour subset of the dataset, was used to train the HMM.
Table \ref{tab:datasets} provides a description of the dataset used for training and testing of the automatic segmentation systems.

For training the ``CleanSpeech'', ``Speech+Music'' and ``Speech+Noise'' classes were treated as a single ``speech'' class.
A two state HMM was defined, with states representing ground truth ``speech'' and ``no-speech'' labels respectively.
The HMM observations are the binary output of the ``speech'' neuron in the CNN, which indicates ``speech'' or ``no-speech''.
Note, the ``no-speech'' label differs slightly from the ``non-speech'' label, since the ``speech'' and ``non-speech'' sounds can co-occur, whereas ``no-speech'' implies ``speech'' does not occur in the signal.

Transition and emission probabilities were trained in a supervised manner, by passing AVA-Speech (train) though the CNN, then using the labels predicted by the CNN and corresponding ground truth labels as observations and hidden state sequences respectively.
Viterbi decoding was then used to find the most likely underlying label sequence, given CNN predicted labels.
Finally the VAD segments are used as input to the X-vector diarization system.

\begin{table}
\resizebox{\columnwidth}{!}{
\begin{tabular}{@{}lll@{}}
\toprule
\textbf{Dataset}        & \textbf{Size (hours)}         & \textbf{Use}          \\ \midrule
Audio Set               & $\approx$~53                  & CNN training          \\
AVA-Speech (train)      & $\approx$~23                  & HMM/GMM-HMM training  \\
AVA-Speech (test)       & $\approx$~23                  & VAD testing           \\\bottomrule
\end{tabular}
}
\caption{Datasets used for training and testing of automatic segmentation systems.}
\label{tab:datasets}
\end{table}

\subsection{VAD${_3}$: CNN-GMM-HMM}
While the CNN-HMM approach yields a large improvement over the energy-based VAD, it may be possible to improve it further by making use of the CNN soft label outputs, rather than the hard labels obtained by taking the argmax of the CNN outputs.
In this case the HMM observation sequence is chosen to consist of the output probabilities computed by the CNN speech neuron, rather than the binary labels.
Where the observations were previously modelled as repeated Bernoulli trials, they are now continuous and can therefore be modelled by a more complex distribution function.
A 3-mixture GMM for each of the two HMM states was found to be an effective choice.
Fewer mixtures led to deteriorated performance, while more mixtures did not result in further improvement.
As before, the GMM-HMM is trained on AVA-Speech (train) in a supervised manner and the resulting segments used as input for the X-vector diarization system.

\begin{figure*}[t]
	\centering
	\includegraphics[width=0.6\textwidth]{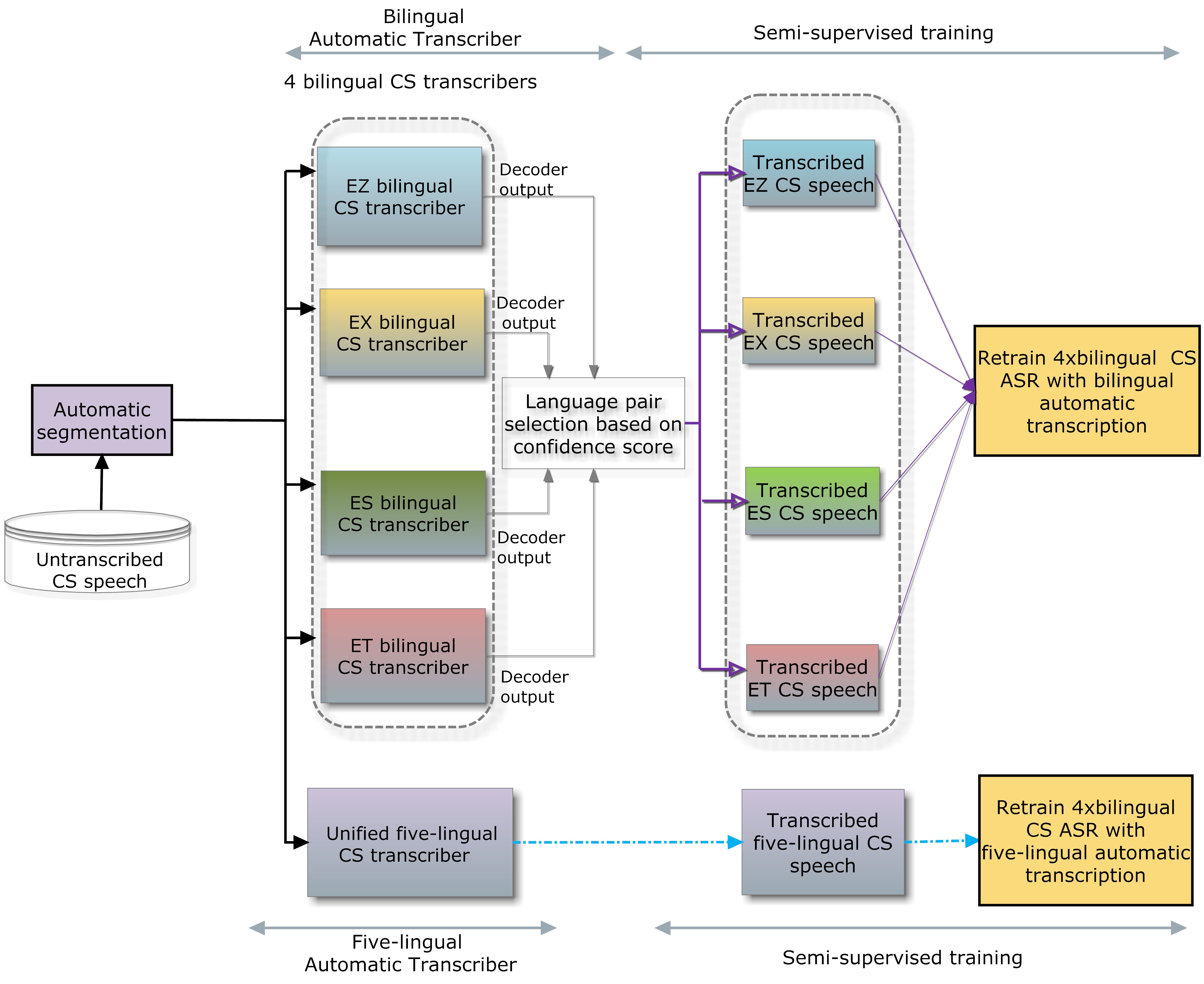}
	\caption{Semi-supervised training framework for bilingual code-switch (CS) ASR. EZ, EX, ES and ET refer to Engish-isiZulu, English-isiXhosa, English-Sesotho and English-Setswana language pairs respectively.}
	\label{mdnn}
	\vspace{-6pt}
\end{figure*}

\subsection{VAD${_4}$: VAD${_3}$ without speaker diarization}
While the X-vector diarization system is useful for obtaining speaker labels for each segment, it is computationally expensive and represents only a pre-processing step for downstream ASR.
To determine its importance, our final experiment used the segments produced by our best performing VAD system directly, without diarization.
Hence each segment was treated as being from a different speaker.
%Results for this experiment, as well as the other automatic segmentation techniques when used for ASR, can be found in Section \ref{ssec:subhead}.

% -----------------------------------------
\section{Automatic Transcription}
\label{sec:auto_trans}
% -----------------------------------------
Recent studies demonstrated that semi-supervised training can improve the performance of Frisian-Dutch code-switched ASR~\cite{yilmaz2018semi} as well as South African code-switched ASR \cite{biswas2019IS2}.
The approach taken in this study is illustrated in Figure~\ref{mdnn}. 
The figure shows the two phases of semi-supervised training for the parallel bilingual as well as five-lingual configurations: automatic transcription followed by bilingual semi-supervised acoustic model retraining.
The five-lingual system was not retrained with the automatically transcribed data for this set of experiments as our primary motive was to study the effect of automatic segmented speech on bi-lingual semi-supervised ASR.
%\trn{can we say why not?}

\subsection{Parallel bilingual transcription}
This system (System A) consists of four subsystems, each corresponding to a language pair for which code-switching occurs. 
Acoustic models were trained on the manually segmented and transcribed soap opera data (ManT, described in Section~\ref{SEC:corpus:transcribed}) pooled with the manually segmented but automatically transcribed speech (AutoT$_{\rm Exp}$, introduced in Section~\ref{SEC:corpus:untranscribed1}).
Because the languages spoken in the untranscribed data were unknown, each segment was decoded in parallel by each of the bilingual decoders.
The output with the highest confidence score provided both the transcription and a language pair label for each segment.

\subsection{Five-lingual transcription}
Some of our previous experiments indicated that the automatic transcriptions generated by the five-lingual baseline model enhanced the performance of the bilingual semi-supervised systems~\cite{biswas2019IS2}. 
%Thus, here we did consider automatic transcriptions by five-lingual baseline system to develop semi-supervised acoustic model.
Five-lingual transcriptions were therefore also included in this study.
The five-lingual system (System H) is based on a single acoustic model trained on all five languages.
%The training data consisted of the manually transcribed soap opera speech (ManT) pooled with the transcriptions generated by a five-lingual system (AutoT\_F) introduced in Section~\secref{SEC:corpus:untranscribed1}.
It was trained on the same data as the bilingual systems, except for the fact that the AutoT$_{\rm Exp}$ data was transcribed using a five-lingual baseline model.

Since the five-lingual system is not restricted to bilingual output, its output allows Bantu-to-Bantu language switching. 
Examples of such switches were indeed observed in the transcriptions.
Moreover, the automatically generated transcriptions sometimes contained more than two languages.
Although the use of more than two languages within a single segment is not common, we have observed such cases during the compilation of the manually transcribed dataset.
For our fast, continuous speech, the automatically generated segments have been observed to produce longer segments than manual segmentation of the data.
This increases the likelihood of multiple language switches within the segment.
Unfortunately, since the automatic segments are generated from untranscribed data, the degree to which multiple languages occur within a single automatic segment is difficult to quantify.

\vspace{-3pt}
% -----------------------------------------
\section{Automatic Speech Recognition}
\label{sec:asr}
% -----------------------------------------
\subsection{Acoustic modelling}
All acoustic models were trained using the Kaldi ASR toolkit \cite{povey2011kaldi} and the data described in Section~\ref{SEC:corpus}. 
The models were trained on a multilingual dataset that included all the data in Table~\ref{tab:duration_unbalanaced_corpora}.
In addition, three-fold data augmentation~\cite{ko2015audio} was applied prior to feature extraction.
The feature set included standard 40-dimensional MFCCs (no derivatives), 3- dimensional pitch and 100 dimensional i-vectors.

The models were trained with lattice free MMI~\cite{povey2016purely} using the standard Kaldi CNN-TDNN-F~\cite{povey2018} Librispeech recipe (6 CNN layers and 10 time-delay layers followed by a rank reduction layer) and the default hyperparameters. 
All acoustic models consist of a single shared softmax layer for all languages, as in general there is more than one target language in a segment.

No phone merging was performed between languages and the acoustic models were all language dependent. For the bilingual experiments, the multilingual acoustic models were adapted to each of the four target language pairs.

%For neural network training, MFCC features were converted to the mel-filter bank features were processed and converted to mel-filter bank features and applied to the input of CNN layer.
%\todo{Ewald: I removed BNF text here.}
%Semi-supervised acoustic model trained with bottleneck features consists of additional 39-dimensional mBNF features. 

\subsection{Language modelling}
\label{sec:typestyle}
%\subsection{Baseline language model}
The EZ, EX, ES, ET vocabularies contained 11\,292, 8\,805, 4\,233, 4\,957 word types respectively and were closed with respect to the training, development and test sets.
The vocabularies were closed since the small datasets and the agglutinative character of the Bantu languages would otherwise lead to very high out-of-vocabulary rates.
The SRILM toolkit \cite{stolcke2002srilm} was used to train and evaluate all language models (LMs). 

Transcriptions of the balanced subset of the ManT dataset as well as monolingual English and Bantu out-of-domain text were used to develop trigram language models.
Four bilingual and one five-lingual trigram language model were used for the transcription systems as well as for semi-supervised training~\cite{yilmaz2018semi,biswas2019IS2}. 
Table~\ref{perplexity} summarises the development and test set perplexities for the bilingual LMs.
Details on the monolingual and code-switch perplexities are only provided for the test set (columns 3 to 6 in Table~\ref{perplexity}).
The test set perplexities of the five-lingual LM are 1007.1, 1881.8, 345.3, and 277.5 for EZ, EX, ES and ET respectively.
Further details regarding the five-lingual perplexities can be found in \cite{biswas2019IS2}.

Much more monolingual English text was available for language model development than text in the Bantu languages (471M vs 8M words).
Therefore, the monolingual perplexity (MPP) is much higher for the Bantu languages than for English for each language pair.

Code-switch perplexities (CPP) for language switches indicate the uncertainty of the first word following a language switch. 
EB corresponds to switches from English to a Bantu language and BE indicates a switch in the other direction. 
Table~\ref{perplexity} shows that the CPP for switching from English to isiZulu and isiXhosa is much higher than switching from these languages to English. 
This can be ascribed to the much larger isiZulu and isiXhosa vocabularies, which are, in turn, due to the high degree of agglutination and the use of conjunctive orthography in these languages.
The CPP for switching from English to Sesotho and Setswana is found to be lower than switching from those languages to English.
We believe that this difference is due to the much larger English training set.
The CPP values are even higher for the five-lingual language model.
This is because the five-lingual trigrams allow language switches not permitted by the bilingual models.

\begin{table*}[t]
\footnotesize
% \fontsize{5.6pt}{7pt}\selectfont
\centering
\begin{tabular*}{\textwidth}{@{\extracolsep{\fill}} l @{\hspace{4pt}} r @{\hspace{4pt}} r @{\hspace{4pt}} r @{\hspace{4pt}} r @{\hspace{4pt}} r @{\hspace{4pt}} r @{\hspace{4pt}} r @{\hspace{4pt}} r @{} }
\toprule
 & Dev & Test & all CPP & CPP$_{\rm EB}$ & CPP$_{\rm BE}$ & all MPP & MPP$_{\rm E}$ & MPP$_{\rm B}$ \\
\midrule
%\multicolumn{9}{c}{\textbf{Bilingual trigram language model}} \\
EZ & 425.8 & 601.7 & 3\,291.9 & 3\,835.0 & 2\,865.4 & 358.1 & 121.1 & 777.8 \\
EX & 352.9 & 788.8 & 4\,914.4 & 6\,549.6 & 3\,785.6 & 459.0 & 96.8 & 1\,355.6 \\ 
ES & 151.5 & 180.5 & 959.0 & 208.6 & 4\,059.1 & 121.2 & 126.9 & 117.8 \\ 
ET & 213.3 & 224.5 & 70.2 & 317.3 & 3\,798.1 & 160.4 & 142.1 & 176.1 \\
\bottomrule
\end{tabular*}%
\caption{Development and test set perplexities. CPP: code-switch perplexity. MPP: monolingual perplexity.}
\label{perplexity}
\end{table*}

\section{Semi-supervised Training}
For semi-supervised ASR, lattice-based supervision was combined with the lattice-free MMI objective function~\cite{manohar2018semi,carmantini2019untranscribed}. 
Conventionally, semi-supervised training only considers the best path while lattice-based supervision uses the entire decoding lattice. 
Hence, the latter approach allows the model to learn from alternative hypotheses when the best path is not accurate.

Table~\ref{tab:system_config} gives an overview of the bilingual ASR systems that were trained using the manually segmented data (System B) as well as five different versions of the automatically segmented data (Systems C-G \& I). 
In addition to manually-transcribed speech, ManT, the AutoT$_{\rm Exp}$ data was also included in all the training sets.

%Speaker diarization was not applied in VAD$_{4}$.
%Each segment was therefore considered to belong to a unique speaker.

\begin{table}[h]
\Huge
\resizebox{0.49\textwidth}{!}{%
%\begin{tabular*}{0.48\textwidth}{@{\extracolsep{\fill}} llcccccc@{}}
\begin{tabular}{@{} llcccccc@{}}
\toprule
&  &  \multicolumn{6}{c}{\textbf{Training segments}} \\
System & Type & \multicolumn{1}{c}{\begin{tabular}[c]{@{}c@{}}AutoT$_{\rm NonE}$\\   (23h)\end{tabular}} & \multicolumn{1}{c}{\begin{tabular}[c]{@{}c@{}}VAD$_1$\\ (83.6h)\end{tabular}} & \multicolumn{1}{c}{\begin{tabular}[c]{@{}c@{}}VAD$_2$\\ (47h)\end{tabular}} & \multicolumn{1}{c}{\begin{tabular}[c]{@{}c@{}}VAD$_{\rm 2Sub}$\\ (20.9h)\end{tabular}} & \multicolumn{1}{c}{\begin{tabular}[c]{@{}c@{}}VAD$_3$\\ (37.0h)\end{tabular}} & \multicolumn{1}{c}{\begin{tabular}[c]{@{}c@{}}VAD$_4$\\ (45.63h)\end{tabular}} \\ \midrule
A & Bilingual baseline  &  &  &  &  &  &  \\ \cmidrule(l){2-8}
B & \multirow{6}{*}{\begin{tabular}[c]{@{}l@{}}Bilingual system\\ trained with AutoT$_{\rm A}$\end{tabular}} & \checkmark &  &  &  &  &  \\ \cmidrule(l){3-8} 
C &  &  & \checkmark &  &  &  &  \\ \cmidrule(l){3-8} 
D &  &  &  & \checkmark &  &  &  \\ \cmidrule(l){3-8} 
E &  &  &  &  & \checkmark &  &  \\ \cmidrule(l){3-8} 
F &  &  &  &  &  & \checkmark &  \\ \cmidrule(l){3-8} 
G &  &  &  &  &  &  & \checkmark \\ \cmidrule(l){2-8}
H & Five-lingual baseline  &  &  &  &  &  &  \\ \cmidrule(l){2-8}
I & \begin{tabular}[c]{@{}l@{}}Bilingual system\\ trained with AutoT$_{\rm H}$\end{tabular} & \multicolumn{1}{l}{} & \multicolumn{1}{l}{} & \multicolumn{1}{l}{} & \multicolumn{1}{l}{} & \multicolumn{1}{c}{\checkmark} & \multicolumn{1}{l}{} \\ \bottomrule
\end{tabular}%
}
\vspace{-6pt}
\caption{ASR systems trained on different versions of the automatically segmented data. The duration of each of these datasets is given in parentheses.} 
%\ewald{At `E' there is `Standard' under the acoustic features? Is this correct and what is it?}}
\label{tab:system_config}
\end{table}

Also defined in Table~\ref{tab:system_config} are systems A and H, which are the bilingual and five-lingual baseline systems respectively, trained only on the ManT and AutoT$_{\rm Exp}$ data.
These baseline systems were used to obtain automatic transcriptions, AutoT$_{\rm A}$ and AutoT$_{\rm H}$, for each version of the additional data shown in Table~\ref{tab:system_config}.
These automatic transcriptions were subsequently used to train new acoustic models.

VAD$_{\rm 2Sub}$ was included to enable a fair comparison between automatic and manual segmentation. 
This is a 21-hour, randomly selected subset of the VAD$_2$ data which is comparable in size to the manually-segmented dataset (AutoT$_{\rm NonE}$). 

% -----------------------------------------
\section{Results \& Discussion}
\label{sec:results}
% -----------------------------------------
The next three subsections concern results of the systems described in Sections \ref{sec:auto_seg}, \ref{sec:auto_trans} and \ref{sec:asr}
Finally, ASR results are presented for specific languages, as well as at code-switching points.
\subsection{Automatic segmentation}
AVA-Speech (test), which is the $\approx$~23 hour subset of AVA-Speech not used for HMM training, was used as a test set to evaluate VAD performance.
This dataset provides similar conditions to our target domain of soap opera data, as well as dense voice activity labels.
Furthermore, it is accompanied by baseline results for the WebRTC project VAD~\cite{WebRTC} as well as two CNN-based systems based on the architecture proposed in~\cite{HersheyCNNVAD}.
The smaller of these two CNN-based systems, \textit{tiny320}, is similar in size to our CNN, also containing three convolutional layers, while the other, \textit{resnet960}, is based on the much larger ResNet-50 architecture~\cite{resnet50}.

Frame-based true positive rates (TPR) for a fixed false positive rate (FPR), scored over 10ms frames are shown for all VAD systems in Table~\ref{tab:VADres}.
To allow comparison, all VAD systems were tuned to achieve a FPR of 0.315, as described in~\cite{ChaudhuriAVA}.
TPR is reported for each individual speech condition (clean speech, speech with noise and speech with music) as well as for all conditions combined.

As expected, VAD${_1}$ performs poorly.
However, it is interesting to note that it is the only system that performs better for the ``Noise'' and ``Music'' conditions than for the ``Clean'' condition.
This is because noisy signals tend to have more energy than their clean counterparts, making noisy signals more likely to exceed an energy threshold.

A large performance improvement is seen for VAD$_2$ which uses the CNN-HMM.
In particular, this system already outperforms~\textit{tiny320}.
A smaller performance increase is reported for VAD$_3$ which uses the CNN-GMM-HMM.
However, this increase brings its performance to a level comparable to the much larger \textit{resnet960} system.
In the case of ``Music'', VAD$_3$ outperforms \textit{resnet960}, whilst for ``All'' the TPR of VAD$_3$ is within 1\% absolute.

In terms of computational complexity, the energy VAD is about 30 times faster than the CNN based VADs. However, the speaker diarization system is two orders of magnitude slower than the slowest VAD, making the compute times of VAD$_1$, VAD$_2$ and VAD$_3$ are roughly equivalent. VAD$_4$, which removes the speaker diarization, is much faster.

%To compare computational efficiency, VAD$_1$, VAD$_2$ and VAD$_3$ were tested on a workstation with a 3.50GHz Intel Core i7-4771 CPU, 32GB of RAM, and inference preformed in single thread mode.
%The average real-time factors (RTF) for VAD$_1$, VAD$_2$ and VAD$_3$ without speaker diarization were $28.602 \times 10^{-6}$, $942.50 \times 10^{-6}$ and $961.62 \times 10^{-6}$, respectively. 
%Speaker diarization, common to all segmentation systems except VAD$_4$ (which is VAD$_3$ without diarization) has a RTF of 
%As expected VAD$_2$ and VAD$_3$ are significantly slower than VAD$_1$, however they are still fast systems, which can be further sped up by multithreading and GPU use.

\begin{table}
\footnotesize
%\resizebox{\columnwidth}{!}{
\begin{tabular*}{0.47\textwidth}{@{\extracolsep{\fill}}lcccc@{}}
\toprule
                & \multicolumn{4}{c}{True positive rate} \\
Model           & Clean     & Noise	    & Music	    & All       \\ \midrule
RTCvad          & 0.786     & 0.706	    & 0.733	    & 0.722     \\
tiny320         & 0.965	    & 0.826	    & 0.623	    & 0.810     \\
resnet960       & 0.992     & 0.944	    & 0.787	    & 0.917     \\ \midrule
VAD${_1}$       & 0.564	    & 0.662	    & 0.693	    & 0.646     \\ 
VAD${_2}$       & 0.972	    & 0.898	    & 0.778	    & 0.886     \\
VAD${_3}$       & 0.985     & 0.917	    & 0.811     & 0.907     \\
\bottomrule
\end{tabular*}
%}
\caption{The true positive rate reported at a false positive rate of 0.315 for various VAD systems tested on AVA-Speech.
The first three systems are baselines from \protect\cite{ChaudhuriAVA}, tested on the full dataset.
The final three systems are tested on a $\approx$~23 hour test set split.}
\label{tab:VADres}
\end{table}

\subsection{Automatic transcription}

\begin{table*}[h!]
\footnotesize
\renewcommand*{\arraystretch}{1.0}
%\resizebox{\columnwidth}{!}{%
\begin{tabular*}{\textwidth}{@{\extracolsep{\fill}}lrrrrrrr@{}}
\toprule
%\multirow{2}{*}{Language} &
System&\multicolumn{6}{c}{A: Bilingual}                                         & H: Five-lingual         \\ \cmidrule(l){2-7} \cmidrule(l){8-8} 
                          Language& \multicolumn{1}{c}{AutoT$_{\rm NonE}$} & \multicolumn{1}{c}{VAD${_1}$} & \multicolumn{1}{c}{VAD${_2}$} & \multicolumn{1}{c}{VAD$_{\rm 2Sub}$} & \multicolumn{1}{c}{VAD${_3}$} & \multicolumn{1}{c}{VAD${_4}$} & \multicolumn{1}{c}{VAD${_3}$} \\ \midrule
English                   & 8\,570      & 12
\,155 & 7\,608  & 4\,721     & 11\,686 & 4\,754  & 23\,973 \\
IsiZulu                   & 5\,955      & 4\,084  & 3\,583  & 2\,065     & 7\,995  & 2\,122  & 7\,315                  \\
IsiXhosa                  & 302                        & 154                    & 116                    & 57                        & 443                    & 236                    & 831                    \\
Sesotho                   & 1\,317      & 2\,267  & 1\,695  & 759                       & 3\,457  & 719                    & 1\,942             \\
Setswana                  & 2\,598      & 6\,272  & 4\,341  & 2\,241     & 6\,691  & 2\,196  & 1\,973             \\
Code-switched             & 25\,824     & 39\,562 & 30\,904 & 12\,572    & 36\,475 & 17\,911 & 30\,616            \\ \bottomrule
\end{tabular*}%
%}
\caption{Number of segments per language  identified by the baseline bilingual (A) and baseline five-lingual (H) ASR systems for different segmentation approaches.}
\vspace{-6pt}
\label{tab:Trans_details}
\end{table*}

The automatic transcription outputs of the bilingual (System A) and five-lingual (System H) baseline systems are summarised in Table \ref{tab:Trans_details}. 
The first five rows of the table correspond to segments that were classified as monolingual while the last row shows the number of segments that contain code-switching.
The values in this row reveal a high number of code-switched segments in the additional data.
%\trn{higher than what?}

In terms of the number of segments per category, the output of the automatic segmentation systems agree with the manual segmentation process. 
The only exception is the number of English segments identified by the five-lingual system, which is higher than for the other systems.
%The five-lingual system were more confident of recognizing English than Bantu languages as
We believe that this is because the five-lingual language model was trained on more in-domain English text~\cite{biswas2019IS2}.

%The table also shows that including speaker diarization in the segmentation process produces smaller chunks of words than manual segmentation.
The table also shows that including speaker diarization in the segmentation process produces smaller chunks of words than using only the VAD.
%In addition, VAD followed by speaker diarization tends to contain more code-switches than the manual segmentation.
Due to the varying duration of each set, comparisons are difficult to make.
For this reason, the set VAD$_{\rm 2Sub}$ is included, which is of a similar duration to AutoT$_{\rm NonE}$, allowing comparison between the non-expert manual segmentation and automatic segmentation.
It can be seen that for the same duration of data, the automatic segmentation produces fewer segments.

%According to the manual segments isiXhosa has the lowest number of monolingual segments. This observation is confirmed by the trend in the results for automatic segmentation. The same can be said for other languages also. 
%The five-lingual transcription system were more confident of recognizing English than Bantu languages as five-lingual language model were trained with larger amount of in-domain English text compared to the bilingual LMs\cite{biswas2019IS2}.

\subsection{Automatic speech recognition}
The performance of the ASR systems introduced in Table~\ref{tab:system_config} were measured in terms of the word error rate (WER) achieved after semi-supervised training.
Results for the different training configurations are reported in Table~\ref{tab:result1}.
The values in the table indicate that including the additional 23 hours of non-expert manually segmented data (AutoT$_{\rm NonE}$) in the training set (System B) yields absolute improvements of 1.5\% and 1.4\% over the baseline (System A) for the development and test sets respectively.
 
\begin{table*}[!ht]
\scriptsize
\renewcommand*{\arraystretch}{1}
%\resizebox{\textwidth}{!}{%
\begin{tabular*}{\textwidth}{@{\extracolsep{\fill}}lllllllllllllllllll@{}}
\toprule
\multirow{2}{*}{CS Pair} & \multicolumn{2}{c}{\textbf{A (baseline)}} & \multicolumn{2}{c}{\textbf{B}} & \multicolumn{2}{c}{\textbf{C}} & \multicolumn{2}{c}{\textbf{D}} & \multicolumn{2}{c}{\textbf{E}} & \multicolumn{2}{c}{\textbf{F}} & \multicolumn{2}{c}{\textbf{G}} & \multicolumn{2}{c}{\textbf{H}} &
\multicolumn{2}{c}{\textbf{I}}\\ \cmidrule(l){2-3} \cmidrule(l){4-5} \cmidrule(l){6-7} \cmidrule(l){8-9} \cmidrule(l){10-11} \cmidrule(l){12-13} \cmidrule(l){14-15} \cmidrule(l){16-17} \cmidrule(l){18-19}
                         & Dev               & Test              & Dev            & Test          & Dev            & Test          & Dev            & Test          & Dev            & Test          & Dev            & Test          & Dev            & Test          & Dev            & Test     & Dev & Test     \\ \midrule
EZ                       & 34.5              & 40.8              & 33.1           & 39.6          & 33.3           & 39.2          & 33.1           & 39.0          & 33.8           & 39.1          & 32.7           & 38.6          & 33.3           & 38.9          &    37.6      &  43.6 & 33.2 & 38.5        \\
EX                       & 35.8              & 42.7              & 35.3           & 42.0          & 34.8           & 41.9          & 34.7           & 41.8          & 35.2           & 42.1          & 34.7           & 41.4          & 34.7           & 42.3          &  40.6      & 54.5  & 33.8 & 41.0      \\
ES                       & 51.7              & 48.7              & 48.7           & 46.5          & 49.8           & 47.3          & 48.7           & 47.7          & 49.0           & 47.1          & 49.2           & 46.8          & 49.1           & 47.9          &  54.5         & 49.3 & 49.6 & 45.7          \\
ET                       & 44.3              & 41.3              & 42.7           & 39.7          & 41.1           & 38.9          & 40.7           & 38.5          & 41.7           & 40.0          & 40.1           & 38.7          & 40.8           & 39.3          & 47.2          &43.9 & 39.9 & 38.4       \\
\textbf{Overall}         & 41.5              & \textbf{43.4}              & 40.0           & \textbf{42.0}          & 39.8           & \textbf{41.8}          & 39.3           & \textbf{41.7}          & 39.9           & \textbf{42.1}          & 39.2           & \textbf{41.4}          & 39.5           & \textbf{42.1}          &  46.5        & \textbf{46.7} & 39.1 & \textbf{40.9}
\\ \bottomrule
\end{tabular*}%
%}
\caption{Mixed WERs (\%) for the four code-switched language pairs.}
\label{tab:result1}
\end{table*}

\begin{table*}[!h]
\footnotesize
%\resizebox{\columnwidth}{!}{%
\begin{tabular*}{\textwidth}{@{\extracolsep{\fill}}lrrcrrcrrcrrc@{}}
\toprule
\multirow{2}{*}{System} & \multicolumn{3}{c}{English-isiZulu}                                      & \multicolumn{3}{c}{English-isiXhosa}                                     & \multicolumn{3}{c}{English-Sesotho}                                      & \multicolumn{3}{c}{English-Setswana}                                     \\ \cmidrule(l){2-4} \cmidrule(l){5-7}  \cmidrule(l){8-10} \cmidrule(l){11-13} 
                        & \multicolumn{1}{c}{E} & \multicolumn{1}{c}{Z} & \multicolumn{1}{c}{Bi$_{CS}$} & \multicolumn{1}{c}{E} & \multicolumn{1}{c}{X} & \multicolumn{1}{c}{Bi$_{CS}$} & \multicolumn{1}{c}{E} & \multicolumn{1}{c}{S} & \multicolumn{1}{c}{Bi$_{CS}$} & \multicolumn{1}{c}{E} & \multicolumn{1}{c}{T} & \multicolumn{1}{c}{Bi$_{CS}$} \\ \midrule
A (baseline)                      & 37.9                  & 48.7                  & 33.3                     & 37.8                  & 54.5                  & 25.8                     & 43.7                  & 61.4                  & 25.2                     & 36.2                  & 51.8                  & 35.6                     \\
B                       & 32.3                  & 45.2                  & 36.8                     & 32.7                  & 49.1                  & 32.1                     & 32.9                  & 57.2                  & 33.7                     & 28.1                  & 48.3                  & 40.5                     \\
F                       & 31.6                  & 43.9                  & 37.8                     & 31.5                  & 48.3                  & 34.2                     & 32.5                  & 56.8                  & 33.8                     & 27.4                  & 46.4                  & 42.2                     \\
%H                       & 31.3                  & 43.7                  & 38.7                     & 32.8                  & 46.8                  & 34.3                     & 32.5                  & 55.8                  & 34.8                     & 27.1                  & 45.8                  & 42.9                     \\
I                     & 31.7                  & 43.7                 & 37.3                    & 31.6                  & 47.6                  & 34.4                    & 32.0                  & 56.4                  & 34.2                    & 26.8                  & 45.7                 & 42.0                     \\
\bottomrule
\end{tabular*}%
%}
\caption{Language specific WER (\%) (lowest is best)  for English (E), isiZulu (Z), isiXhosa (X), Sesotho (S), Setswana (T) and code-switched bigram correct (Bi$_{CS}$) (\%) (highest is best) for the test set.}
\label{tab:results2}
\end{table*}

The results for System C show that using 83 hours of automatically-segmented speech results in an absolute improvement of 1.7\% and 1.6\% for the development and test sets relative to System A. 
Although System C's performance is on par with that of System B, its training set was much larger which means that the computational cost of developing the system is also much higher. 
%The automatic transcription of 83 hours of automatically segmented speech followed by filtering and retraining are not computationally effective. 

According to Table~\ref{tab:system_config}, VAD$_2$ reduced the additional training data from 83.6 to 47 hours.
Furthermore, including this reduced additional data in System D's training set resulted in lower WERs for both the development and test sets, when compared with Systems A, B and~C.
%The CNN-HMM based automatic segmentation technique (VAD$_2$) significantly reduced the amount of automatically segmented speech than VAD$_1$ but the semi-supervised system (D) trained with VAD$_2$ outperforms semi-supervised systems A, B and C. 
%In contrast with VAD$_1$ which contains 83.6h of speech, VAD$_2$ contains 47.06h of automatically transcribed speech and yet it is capable of showing enhanced performance compared to B. 

The semi-supervised system trained on the 21-hour subset of VAD$_2$ (System E) achieved results that are comparable to those of System B. 
The two additional dataset seem to have had almost the same impact on the accuracy of the resulting acoustic models.
%The approximately same amount of automatically segmented and manually segmented data had the similar impact on semi-supervised acoustic model.
This result seems to indicate that, in terms of ASR performance, manually and automatically-produced segmentations are equally well suited for system development.
%From this observation, it can be said that the quality of automatic segmentation is good as manual segmentation. Although it should be kept in mind that manual segmentation were carried by volunteers not by language experts.  
However, it should be kept in mind that the segment labels used by System B were not assigned by experts.

System F, trained on the segments generated by VAD$_3$, yielded better performance than system D, despite the fact that System D's training set contained 10 more hours of data.
%The system was trained with 10 hour of less automatic segmented speech but yet delivers 1.4\% better performance overall on test set than system D. 
The improvement in WER was found to be statistically significant at the 95\% confidence level using bootstrap interval estimation \cite{bisani2004bootstrap}. 

The results for System G show that automatic segments that do not take speaker identity into account (VAD$_4$) do not achieve the accuracy levels as those that do (System F).
Therefore, the inclusion of speaker diarization does tend to improve ASR performance.
%had failed to capture speaker specific information that yields performance drop compared to F as expected.  

The performance of System H (five-lingual baseline system) is included in Table~\ref{tab:result1} but should not be directly compared with the bilingual systems because the recognition task is inherently more complex.
%it would be unfair to compare it with the bilingual systems because the
%Because five-lingual recognition is more difficult since it allows more freedom in terms of the permissible language switches 
However, as has been observed before~\cite{biswas2019IS2}, the bilingual System I, trained on automatic transcriptions generated by System H, shows the best overall performance of all the evaluated systems. 
The improvement on the test set over its closest competitor (System F) is 0.5\% absolute and this was found to be statistically significant above the 90\% confidence level using bootstrap interval estimation. 
This improvement may be due to the ability of the five-lingual system to transcribe more than two languages, as well as Bantu-to-Bantu switches.
The untranscribed soap opera speech is known to contain at least some segments that do not conform to the four considered bilingual language groupings.
The degree to which such language switches do occur is unfortunately difficult to quantify without manual transcriptions.
However, since the key difference between the bilingual and five-lingual systems is the ability to handle a greater variety of language switches, we speculate that this is a likely cause for the superior performance of System I.

%The semi-supervised system (H) trained with V3 automatic segments with additional multilingual 39 dimensional BNF features exhibits best performance among all. The improvements on development and test set are 0.4\% and 0.5\% absolute respectively compared to the closest competitor F. BNF features helps to reduce the WER of English-Sesotho CS pair by significant margin. The overall improvement over system F was found to be statistically significant at more than 90\% confidence level using bootstrap interval estimation. 

\begin{figure} [t]
	\centering
	\includegraphics[width=\columnwidth]{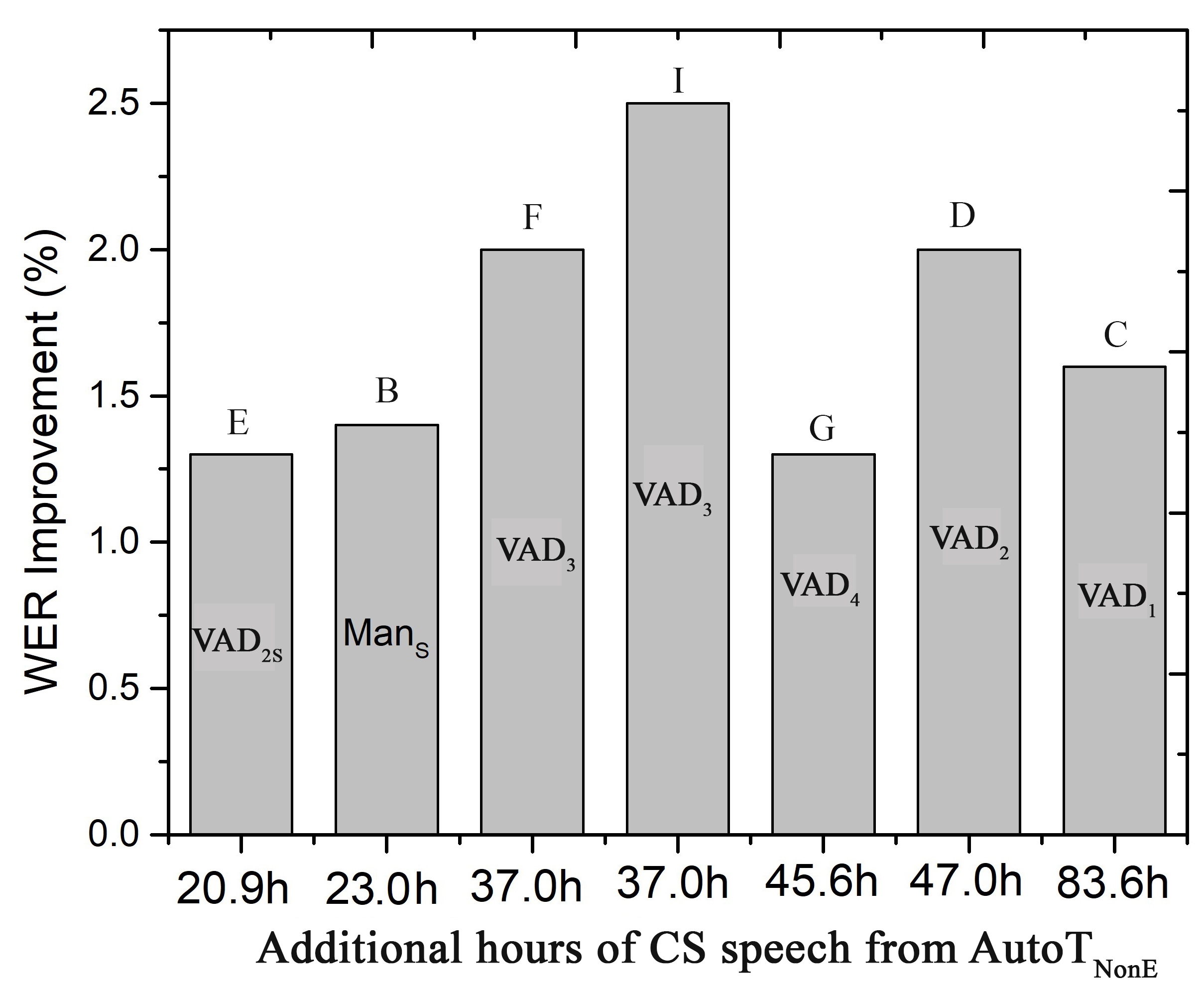}
	\vspace{-6pt}
	\caption{Improvement (\% in comparison with the baseline) in test set WER for different semi-supervised systems incorporating additional soap opera training data.}
	\label{fig:improvement}
		\vspace{-6pt}
\end{figure}

The improvement in WER achieved by the semi-supervised ASR systems incorporating different versions of the additional data is summarised in Figure~\ref{fig:improvement}. 
The figure confirms that the largest gain in recognition accuracy was achieved by System I.
It also affirms the observation that an equal amount of manually and automatically segmented data yields an equal improvement in recognition accuracy in a semi-supervised set-up.

\vspace{-3pt}
\subsection{Language specific WER analysis}
\label{subsec:lan_specific_wer}
For code-switched ASR, the performance of the recogniser at code-switch points is of particular interest. 
Language specific WERs and code-switched bigram correct (Bi$_{CS}$) values for the different semi-supervised systems are presented in Table \ref{tab:results2}. Code-switch bigram correct is defined as the percentage of words correctly recognised immediately after code-switch points.
All values are percentages.

The table reveals that both the English and Bantu WERs for all the semi-supervised systems are substantially lower than the corresponding values for the baseline system.
%all semi-supervised system helped to improve the English and Bantu WER by a significant margin over the baseline system A. 
%It is also interesting to note that multilingual BNF features contributed significantly to reduce the Bantu WER specially for isiXhosa, Sesotho, Setswana but failed to contribute on English WER. It is obvious because multilingual BNF extractor was trained on 9 Bantu languages and BNF features were expected to learn more about Bantu languages. Thus inclusion of English in BNF extractor could lead to improve the performance of English and would be a part of future study. 
The accuracy at the code-switch points is also substantially higher for the semi-supervised systems.
%than baseline system that yields the automatically segmented speech is capable to enhance the performance at code-switch points.
Hence, adding the additional training data enhances system performance at code-switch points.
Moreover, there are no substantial differences between the gains achieved by adding the manually (System B) or automatically (Systems F, I) segmented data.

% -----------------------------------------
\section{Conclusions}
% -----------------------------------------
%\trn{please check ... have I got it right?}
In this study, we have evaluated the impact of using automatically-segmented instead of manually-segmented speech data for semi-supervised training of a code-switched automatic speech recognition system.
%Four different English-Bantu CS pairs were studied.
%The ASR systems were trained in a semi-supervised manner using segments generated by humans as well as automatic segmentation techniques. 
%Almost 250 multilingual soap opera episodes were segmented and transcribed automatically followed by lattice supervised LF-MMI s-supervised training.
Four different automatic segmentation approaches were evaluated, based respectively on simple energy thresholding with diarization, a CNN classification with two variants of HMM smoothing and diarization, and CNN classification with GMM-HMM smoothing and no diarization.
It was found that applying our new CNN-GMM-HMM based VAD followed by X-vector speaker diarization resulted in the best ASR performance.
The results also showed that the performance of systems that used automatically and manually-segmented data were comparable. 
%ed to ensure the quality of automatic transcriptions.
We conclude that automatic-segmentation in combination with semi-supervised training is a viable approach to enhancing the recognition accuracy of a challenging five-language code-switched speech recognition task.  
This is a very positive outcome, since the difficulty in providing a manual segmentation of new broadcast material has remained an impediment to the development of speech technology in severely under resourced settings such as the one we describe.
%In this paper, we also introduce our first approach to develop a multilingual bottleneck extractor trained on nine Bantu languages. Preliminary experiments shows that the multilingual bottleneck features were found to have positive impact on the code-switched acoustic model.
Future work will focus on improving the VAD and speaker diarization techniques as well as incorporating language identification into the automatic segmentation process.

%Future work will focus on developing more capable automatic segmentation techniques as well as speaker and language diarization to extend the pool of training data. 
%The bottleneck feature extractor also needs to be tuned for our application and the impact of the bottleneck feature dimension on the acoustic model needs to be studied.
%in relatively smaller batches to study further potential improvements.
%
% ----------------------------------------------------------
%
\section{Acknowledgements}
We would like to thank the Department of Arts \& Culture (DAC) of the South African government for funding this research. 
We are grateful to e.tv and Yula Quinn at Rhythm City, as well as the SABC and Human Stark at Generations: The Legacy, for assistance with data compilation. 
We also gratefully acknowledge the support of the NVIDIA corporation for the donation of GPU equipment.

% -------------------------------------------------------------------------
\section{References}

\bibliographystyle{lrec}

%\bibliography{lrec2020W-xample}

%\section{Language Resource References}
%\label{lr:ref}
%\bibliographystylelanguageresource{lrec}
%\bibliographylanguageresource{lrec2020W-xample}

\end{document}